\documentclass[aps,prb,twocolumn%
 amsmath,amssymb,
 aps, physrev,
 twocolumn, prm
]{revtex4-2}

\usepackage{graphicx}
\usepackage{dcolumn}
\usepackage{bm}
\usepackage{xcolor}
\usepackage{hyperref}
\usepackage{comment}
\usepackage{amsmath}

\begin{document}

\preprint{APS/123-QED}

\title{\textbf{Improved capabilities of the TurboGAP code for radiation induced cascade simulations: an illustration with silicon} 
}%

\author{Uttiyoarnab Saha\textsuperscript{1}}
\author{Ali Hamedani\textsuperscript{1}}
\email{Corresponding author; ali.hamedani@aalto.fi}
\author{Miguel A. Caro\textsuperscript{2}}
\author{Andrea E. Sand\textsuperscript{1}}

\affiliation{}
\affiliation{\textsuperscript{1}Department of Applied Physics, Aalto University, 00076 Aalto Espoo, Finland}
\affiliation{\textsuperscript{2}Department of Chemistry and Materials Science, Aalto University, 02150 Espoo, Finland}

\begin{abstract}
TurboGAP is a software package designed for efficient molecular dynamics simulations using Gaussian Approximation Potential (GAP) machine-learning interatomic potentials (MLIP). In this work, we enhance the capabilities of TurboGAP for radiation damage simulations by implementing a two-temperature molecular dynamics model, based on electron density-dependent coupling of electronic and atomic subsystems. Additionally, we implement adaptive calculation of the timestep and grouping of atoms for cell-border cooling. Our implementation incorporates electronic stopping power either through a traditional friction-based model or a more realistic first-principles-derived model. By combining the computational efficiency of TurboGAP with the accuracy of GAP MLIP, we perform cascade simulations in silicon with primary knock-on atom (PKA) energies up to 10 keV. Our simulations scale to systems containing up to 1 million atoms. We study the generation and clustering of radiation-induced defects. We also calculate ion-beam mixing and compare our results with the experimental data, discussing how the GAP-MLIP along with the inclusion of a realistic electronic stopping model improves the prediction of experimental mixing values.
\end{abstract}

\maketitle


\section{\label{sec:intro}Introduction}%

Atomic-scale simulation is a vital part of the multiscale modeling framework for materials science. When it comes to the accuracy of the atomistic simulations, the representation of the potential energy surface (PES) as a function of atomic positions takes the highest importance. While quantum mechanical calculations, such as density functional theory (DFT), offer accurate and transferable representations of the PES, their high computational cost restricts simulations to system sizes of only a few hundred atoms and timescales of hundreds of picoseconds. The simulation of larger systems or longer time scales is achieved within the molecular dynamics framework using interatomic potentials (IP), where the IP predicts the potential energy and forces acting on the atoms for the given atomic configuration. 

Traditional IPs empirically parameterize the PES based on the physical functional forms that are dependent on the atomic degrees of freedom \cite{muser2023interatomic}. Machine learning interatomic potentials (MLIPs) approximate the PES by sampling the system's configurational space. Most MLIPs estimate energy based on the locality of the atomic interactions \cite{mishin2021machine}, where local energies ($E_\mathrm{i}$) are associated with "descriptors" of the local environment \cite{himanen2020dscribe,bartok2010gaussian} that remain invariant to translation, rotation, and permutation of identical atoms. The overall PES is then constructed by summing these local mappings ($E=\sum_{i}E_\mathrm{i}$). The selection of the descriptor and the design of the dataset mostly depend on the material in hand and the intended application for which the potential is generated \cite{zuo2020performance}.

There are various MLIP frameworks \cite{behler2007generalized,bartok2010gaussian,caro2019optimizing,thompson2015spectral,shapeev2016moment,drautz2019atomic,batatia2022mace}. Along with many other applications, MLIPs have also been employed to investigate radiation-induced damage. As such, they have been used to study defect dynamics and primary radiation damage in Si \cite{hamedani2021primary, hamedani2020insights, niu2023machine}, W \cite{byggmastar2019machine,liu2023large,wei2023effects,dominguez2020classification}, Al \cite{wang2019deep}, Zr \cite{wang2024properties}, SiC \cite{liu2024deep}, and high-entropy alloys (HEA) \cite{byggmastar2021modeling}. Compared to empirical potentials, these studies have shown several considerable differences in the prediction of the defect state. 

The Gaussian approximation potential (GAP) \cite{bartok2010gaussian} is a kernel-based MLIP framework which utilizes the smooth overlap of atomic environment (SOAP) descriptor \cite{bartok2013representing} to encode the atomic environments. GAP potentials have been successfully employed for cascade simulations, as demonstrated in Refs. \cite{hamedani2021primary, hamedani2020insights}, where the highest energy of the primary knock-on (PKA) atom was 2 keV. However, the computational expense of evaluating the SOAP descriptor poses a challenge for cascade simulations involving higher PKA energies. For instance, simulating a PKA energy of 10 keV in Si would require a pristine simulation cell containing at least $10^{6}$ atoms, significantly increasing the computational cost. With respect to the efficiency, the tabGAP \cite{tabGAP} potential—a low-dimensional variant of GAP—is computationally more efficient, but it is primarly suited for metals, and does not perform as the GAP does for covalent materials \cite{gao_junlei}.

The turboSOAP descriptor \cite{caro2019optimizing} is an optimized variant of the SOAP descriptor. Through faster computation of descriptors and smoother radial basis functions \cite{caro2019optimizing, wang2022structure}, turboSOAP provides improved computational efficiency and accuracy over SOAP. Several GAP potentials that are trained with the turboSOAP descriptor are reported in \cite{kloppenburg2023general,muhli2021machine,wang2022structure}. TurboGAP \cite{caro2019turbogapwebsite} is a versatile molecular dynamics (MD) software which is specifically designed to utilize GAP MLIPs. This specialty enables computationally efficient GAP simulations, where both the SOAP and turboSOAP descriptors are supported. This improvement in the computational efficiency opens the way for cascade simulations with GAP with a few tens of keV PKA energies. However, an MD code requires some essential features to be suitable for such a study. The three critical features are 1) the capability to handle the evolution of the atomic trajectories in adaptive time steps \cite{mdrange}, 2) capability of cooling the cell at the border \cite{andrea-cellBorder}, and 3) to account for the dissipation of the energy of atoms through electronic stopping \cite{nordlund2018primary}.

An energetic ion loses its energy mainly by two mechanisms; through exciting the electrons in its path and by elastic interaction with the atoms. The energy lost through the latter process, known as \textit{nuclear stopping}, can be modeled by classical ion-ion interaction. However, the energy loss to the electronic system, \textit{electronic stopping}, would require a higher level of description and modeling \cite{sand2019heavy}. Different methods have been devised to account for electronic effects in MD simulation of radiation damage. Nevertheless, it is quite challenging to describe the physical phenomena in the full energy range of meV-MeV, that is, from energy regimes of dominant electron-phonon (e-ph) interactions to those of dominant electronic stopping (ES).

In MD simulations, the ES has been traditionally and most often accounted for by simply considering a velocity-dependent \textit{friction force} on energetic atoms. However, in the friction based stopping (FES) method, in order to prevent the unphysical quench of thermal modes in MD simulations, a low-end cutoff value should be considered for the kinetic energy (velocity). However, this cutoff value is often set arbitrarily and this can significantly affect the defect microstructure as observed in W, Ni, and Fe \cite{sand2013high,bjorkas2009assessment,sand2015lower}.

The EPH model \cite{caro2019role,tamm2019role,andrea_nickel}, provides a cutoff-free ground to activate both the ES and e-ph coupling regimes. It has been suggested \cite{caro1989ion} that local electron density can serve as a key parameter for constructing a model that unifies both regimes within a single framework. This approach stems from the observation that the coupling strength in the e-ph regime is directly related to atomic electron densities. Additionally, time-dependent density functional theory (TDDFT) \cite{tddft} has demonstrated the capability to predict electronic stopping power (ESP) for realistic materials by simulating electron excitation processes directly \cite{correa2018calculating}. Consequently, the relationship between the energy losses to the electronic system ($E_{\mathrm{ES}}$) and electron density can also be established. The EPH model is incorporated into the generalized two-temperature molecular dynamics (TTMD) framework, where the atomic subsystem is described using IP, and the energy feedback from the electronic subsystem is incorporated through implementation of stochastic forces. The stochastic forces are modeled using generalized Langevin dynamics \cite{tamm2018langevin}, and interpreted to arise from electronic degrees of freedom.

Here, we present the implementation of three modules in TurboGAP, in order to enhance its performance in radiation damage simulations. First, to account for energy losses to the electronic system, we implement two dissipation models, the EPH and FES models. The EPH model is implemented through the TTMD framework, whereas in the FES model the pre-calculated electronic stopping power is introduced as a friction term in conventional MD. The second module is the adaptive calculation of timestep to account for drastic changes in the atom velocities in close collisions. The third module is the grouping of atoms, with which the temperature of the simulation is controlled through the cooling of the cell at its borders. We benchmark these implementations against similar features available in the LAMMPS \cite{LAMMPS} code and demonstrate their computational efficiency on high-performance computing (HPC) setups. Using the implemented dissipation models, we perform cascade simulations with PKA energies of up to 10 keV in silicon. These simulations are carried out with a GAP MLIP, utilizing turboSOAP as the many-body descriptor. The results from the EPH model are compared to those from the FES model, where the effect of the low-end cutoff for the atoms kinetic energy in the FES model is also investigated. We look at the number of surviving defects and their clustering. We calculate the ion beam mixing in silicon and compare our results with experimental data. We discuss the impact of incorporating realistic electronic stopping models on the calculated mixing values.
\section{\label{sec:modules}new modules in TurboGAP}

\subsection{\label{sec:elstop}Electronic Energy Losses}
\subsubsection{\textbf{FES model}}
 With the FES model, the energetic atoms are decelerated by a scalar friction force, with its magnitude calculated based on the ESP. The friction force component is introduced in the equation of motion, as presented in Eq. \ref{Eqn:vdependentFriction} \cite{nordlund1998defect}. Here, the first term on right hand side is the force derived from interatomic potential and the second term is the resistive forces, where $S_e$ is the ESP. For the energy range of interest, $S_e$ can be calculated using an ion transport software such as SRIM \cite{ziegler2013srim}.

\begin{equation}
	\label{Eqn:vdependentFriction}
	\mathbf{F}_I = -\mathbf{\nabla}_IU - S_e\frac{\mathbf{v}_I}{v_I}
\end{equation}

In our implementation in TurboGAP, the FES is activated with the \verb|electronic_stopping| command. In this implementation, the ESP is obtained from a file where it has been tabulated as a function of energy. The kinetic energy and the ESP are stored in the units of eV and eV/\AA, respectively. In the input file, the user provides the cutoff energy value (\verb|eel_cut|). The friction force is applied on the atoms whose kinetic energy is higher than the cutoff energy. The output data provides 1) $E_{\mathrm{ES}}$ on the current MD step, 2) cumulative $E_{\mathrm{ES}}$ up to the present MD time, 3) kinetic energy and 4) temperature of the group of atoms on which this model is made to act. The output data can be requested at a certain interval of MD steps (\verb|eel_freq_out|).

\subsubsection{\textbf{EPH model}}
With the EPH model the motion of the atoms is described with Eq. \ref{eq:eph} \cite{tamm2018langevin}.

\begin{equation}
	\label{eq:eph}
	\mathbf{F}_I = -\mathbf{\nabla}_IU_{adiab} - \sum_J B_{IJ}\mathbf{v}_J + \sum_J W_{IJ}\mathbf{\xi}_J .
\end{equation}

The first term represents the adiabatic forces that are determined with the IP in hand (in this paper an MLIP has been used in all simulations). The second term describes the friction forces that are linearly dependent to the velocities. The third term represents the random forces carrying the many-body correlations described through tensor combination of atomic coordinates. The tensor $W_{IJ}$ for the random force is defined to be a function of relative atomic positions and the electronic density-dependent coupling parameter, from which the friction tensor $B_{IJ}$ are computed following the fluctuation-dissipation theorem \cite{kubo1966fluctuation}. The random vectors $\xi_J$ are sampled from a normal distribution with the variance of $2k_BT_e$, where $T_e$ is the temperature in the electronic system at the position of atom \textit{I}. The electronic system (a continuum heat bath) is modeled with a heat diffusion equation, Eq. \ref{eqn:hdeqn}, 

\begin{equation}
	\label{eqn:hdeqn}
	C_e(T_e)\frac{\partial T_e}{\partial t} = \mathbf{\nabla}.(\kappa_e(T_e)\mathbf{\nabla}T_e) + Q_{ei} + S_{ext}.
\end{equation}

which is solved simultaneously with the atom dynamics in the process of MD. Here the electronic heat capacity, $C_e$, and conductivity, $\kappa_e$, govern the temperature $T_e$ within the electronic system. The term $Q_{ei}$ handles the local transfer of energy between the atomic and electronic subsystems. For a detailed account of the EPH model and forms of the $W_{IJ}$ and $B_{IJ}$ tensors, see Refs. \cite{tamm2018langevin,caro2019role,tamm2019role}.

In the implementation in TurboGAP, the EPH model is activated using the \verb|nonadiabatic_processes| command. The heat diffusion equation is solved with the finite difference method (FDM) by constructing a suitable mesh depending on the custom-defined dimensions with the number of grid points separated in x (\verb|eph_gsx|), y (\verb|eph_gsy|), and z directions (\verb|eph_gsz|). The FDM time-step (\verb|eph_fdm_steps|), the electronic heat capacity per unit volume (\verb|eph_C_e|), and electronic heat conductivity (\verb|eph_kappa_e|) can also be defined by the user. Moreover, the user has the option for performing simulations with any of the friction forces (\verb|eph_friction_option|), random forces (\verb|eph_random_option|) and the FDM update of $T_e$ (\verb|eph_fdm_option|) activated or deactivated. In addition to the local energy transfer term, external sources, $S_{ext}$, on the electronic subsystem may also be considered. In our implementation, the $C_e$ and $\kappa_e$ can be either independent of or dependent on the temperature. Hence, both the linear and non-linear cases of Eq. \ref{eqn:hdeqn} can be solved. From this calculation the outputs are the energies dissipated due to friction (\verb|E_fric|) and random forces (\verb|E_rand|) on the current MD step, cumulative energy transfer (\verb|E_net_cum|) until the current MD time, electronic temperature (\verb|T_e|), and the atomic temperature (\verb|T_a|) and energies (\verb|Kin.E_a, Pot.E_a|) of the group of atoms (\verb|eph_groupID|) on which the model is made to act. The three dimensional temperature profile in the electronic subsystem is also produced in a separate file (\verb|eph_Toutfile|).

\subsection{\label{sec:adapt_dt}Adaptive timestep}
In the collision cascades, high kinetic energies result in close collisions where atoms experience a steep potential energy gradient. Consequently, to accurately capture the dynamics that is involved in the collisions, the MD timestep must be smaller than that used for the thermalization or near-equilibrium states. Similarly, 
when the cascade cools down and the system goes toward equilibrium, there is no need to use the small timestep that was used in the initial stage of simulation.

To account for this feature, the \textit{adaptive} calculation of the timestep \cite{mdrange} has been implemented in the TurboGAP code. With this module, the timestep is recalculated every $N^{\text{th}}$ steps based on two predetermined criteria of maximum allowed displacement ($x_{\mathrm{max}}$) and change in the kinetic energy ($e_{\mathrm{max}}$) of any atom in the system. In general, the maximum displacement criterion applies throughout the trajectory of an atom, while the time-steps begin to actually sense the maximum energy change criterion when two atoms have moved very close to each other and the potential energy changes sharply in the repulsive interaction regime. In an adaptive-timing simulation using the TurboGAP code, three timesteps are calculated, two from the velocities and accelerations using the maximum displacement criterion, and the third, from the forces using the maximum energy change criterion. Then, the minimum of these three timesteps is selected. Finally, the time step is again scaled such that displacement of any atom with the new time step is less or equal to the set maximum allowable displacement criterion.

In our implementation, the adaptive calculation of the time step is enabled using the \verb|adaptive_time| command. The interval for re-calculating the time step is specified with the \verb|adapt_tstep_interval| command. The values of $x_{\mathrm{max}}$  and $e_{\mathrm{max}}$  are set using the \verb|adapt_xmax| and \verb|adapt_emax| commands, respectively. Finally, the minimum and maximum time step limits are controlled by the \verb|adapt_tmin| and \verb|adapt_tmax| commands, respectively.

\subsection{\label{sec:atom_groups}Grouping of atoms}

In the context of radiation damage simulations, \textit{border cooling} is a common approach in which a thermostat is applied to the atoms located within a certain distance from the cell boundaries. Controlling the temperature at the cell border while initiating a cascade at the center enables the dissipation of the excess heat introduced by the PKA into the bulk of the material. Consequently, effective border cooling ensures that cascade simulations are conducted at the desired temperature.

We present the group-selection capability of atoms in the TurboGAP code, which is enabled using the \verb|make_group| command. Atoms can be selected based on various styles (e.g., \verb|block|, \verb|add|, \verb|atomtype|, etc.), along with inter-group operations and the ability to dynamically update groups based on the positions of their constituent atoms throughout the simulation. For detailed instructions on the selection styles mentioned above, we refer the reader to the documentation at \cite{make_group_manual}.

\section{GAP MLIP}
\label{sec:GAP}

The details about the GAP formalism and turboSOAP descriptors can be found in Refs. \cite{deringer2021gaussian} and \cite{caro2019optimizing}, respectively. For this study, we re-train a Si GAP model with a two-body (2b), a three-body (3b), and a turboSOAP descriptor over the dataset presented in Ref. \cite{bartok2018machine}. The kernel and descriptor hyperparameters are compiled in Table \ref{tab:hyperparams}. For all except liquid and amorphous structures, a uniform regularization of ($\sigma_{\mathrm{energy}}= 1 \;\mathrm{meV/atom}$, $\sigma_{\mathrm{force}}=0.1 \;\mathrm{eV/\AA}$, $\sigma_{\mathrm{stress}}=0.05 \;\mathrm{eV}$) was used in fitting the potential. Regularization parameters for liquid and amorphous structures are $\sigma_{\mathrm{liquid}}=(0.003, 0.15, 0.2)$ and $\sigma_{\mathrm{amorph}}=(0.01, 0.2, 0.4)$ for (energy, force, and stress), respectively.
\begin{table}[ht]
	\centering
	\caption{The hyperparameters used for fitting the GAP MLIP. A detailed description of individual parameters are found in Ref. \cite{klawohn2023gaussian}.}
	\label{tab:hyperparams}
	\begin{ruledtabular}
		\renewcommand{\arraystretch}{1.2} 
		\begin{tabular}{l@{\hskip 8pt}c@{\hskip 8pt}c@{\hskip 8pt}c} 
			& \multicolumn{3}{c}{descriptor}\\ 
			\cline{2-4}
			Hyperparameter& 2b & 3b & turboSOAP \\
			\hline
			$r_{\mathrm{cut}}$ ($\mathrm{\AA}$) & 5.0 & 4.0 & - \\
			$r_{\mathrm{cut}}^{\mathrm{hard}}$ ($\mathrm{\AA}$) &  -	& - & 5.0 \\
			$r_{\mathrm{cut}}^{\mathrm{soft}}$ ($\mathrm{\AA}$) & - & - & 4.0 \\
			\hline
			$l_{\mathrm{max}}$ & -  & -  & 8 \\
			$\alpha_{\mathrm{max}}$ & -  & -  & 8 \\
			$\sigma_{\mathrm{at}}^{r}$ (\AA)& -  & -  & 0.5 \\
			$\sigma_{\mathrm{at}}^{t}$ (\AA) & -  & -  & 0.5 \\
			$\zeta$ & -  & -  & 6 \\
			\hline
			$\delta$ (eV) &  1.0 & 0.01 & 1.0  \\
			$\theta$ & 1.0 & 4.0 & - \\
			\hline
			$N_{\mathrm{sparse}}$ & 25 & 200 & 9000 \\
			Sparse method  & uniform & uniform & CUR \\
		\end{tabular}
	\end{ruledtabular}
\end{table}

We also append the universal Ziegler-Biersack-Littmarck (ZBL) potential \cite{zbl} to capture the repulsion of the energetic atoms involved in the collision cascades.
The ZBL potential is appended as an external potential ($E_{\mathrm{rep}}$) in a tabulated format, where energy is given as a function of atomic distance. In training the GAP model, the tabulated potential is taken as a baseline which is subtracted from the reference energies and is added back in prediction. The ZBL potential is defined as in Eq. \ref{eq:zbl} 
\begin{equation}
	\label{eq:zbl}
	E_{\mathrm{rep}}(r) = \frac{Z_1 Z_2 e^2}{4\pi \epsilon_0 r} \phi(r/a) f_{\mathrm{cut}}(r)
\end{equation}
where the screening function, $\phi(x)$ ($x = r/a$), is given as
\begin{align}
	\label{eq:screening_function}
	\phi(x) &= 0.1818 e^{-3.2x} + 0.5099 e^{-0.9423x} \notag \\
	&\quad + 0.2802 e^{-0.4029x} + 0.02817 e^{-0.2016x}
\end{align}
In Eq. \ref{eq:screening_function}, the screening length ($a$) is defined as
\begin{equation}
	a = \frac{0.46848}{Z_1^{0.23} + Z_2^{0.23}}
\end{equation}
The cutoff function, $f_{\mathrm{cut}}(r)$, is given by
\begin{equation}
f_{\text{cut}}(r) =
\begin{cases}
	1, & r \leq r_1 \\
	1 - \eta^3 (6\eta^2 - 15\eta + 10), & r_1 \leq r \leq r_2 \\
	0, & r \geq r_2
\end{cases}
\end{equation}
with $\eta=(r - r_1)/(r_2 - r_1)$. The cutoff function is used to switch the $E_{\mathrm{rep}}$ off in the near-equilibrium regime, where the dominant part of the reference data reside. Similarly, in the ranges where the high energy from binary interactions dominates ($r \leq r_1$), the $E_{\mathrm{rep}}$ takes over. The $r_1$ and $r_2$, were set to be 1.0 and 2.5 \AA, which were selected by monitoring different values for a smooth transition between the high-energy and near-equilibrium ranges at $r_1$ and $r_2$.

\begin{figure}[ht]
	\centering
	\includegraphics[width = 0.8\columnwidth]{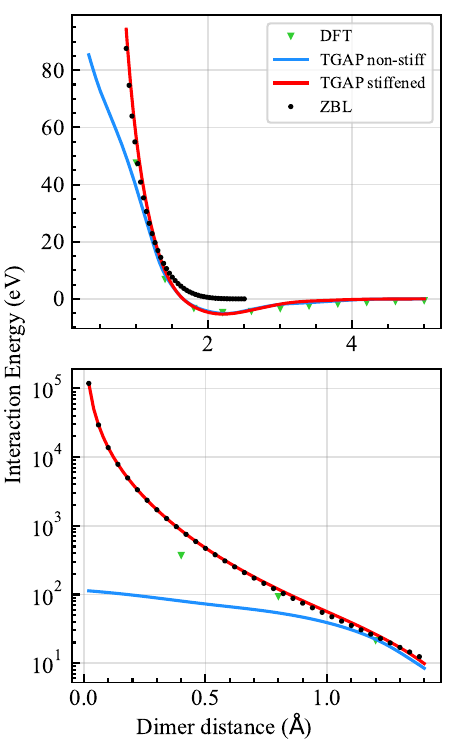}
	\caption{The performance of the trained GAP MLIP, stiffened with the ZBL repulsive potential, in predicting the Si-Si dimer interaction in the given range of distance. The "non-stiff" curve shows the prediction of the potential in the absence of the repulsive potential. The prediction of the forces is presented in Supplementary Fig.1. The details of DFT-dimer calculations is also presented in Supplementary material.}
	\label{fig:dimer}
\end{figure}

Fig. \ref{fig:dimer} demonstrates the performance of the potential in reproducing the interaction of Si-Si dimer. As seen, the GAP MLIP follows the DFT data points
 down to $\approx$ 1 \AA, after which alignment with the ZBL potential starts. The prediction of forces is shown in Supplementary Fig. 1.

\section{Speed and scaling on HPC setups}
\label{bsec:speed-scaling-HPC}

The introduced modules have been implemented with the open message passing interface (OpenMPI) parallelism in the TurboGAP code. It is to be noted that the computation of forces in the EPH model, in particular, requires parallelization for efficient simulations. On the other hand, while using the adaptive timestep and ES model only, the efficiency due to parallelization will be more evident only in case of simulating  really large systems. We study the scaling of the implemented modules using multiple processors. A fixed number (500) of MD steps are performed for four representative modes, viz. only EPH model (called M2), only adaptive timestep (M3), adaptive timestep with FES model (M4), and adaptive timestep with EPH model (M5). Two simulation boxes containing 8000 and $10^6$ atoms were used in the tests. All simulations were performed employing the potential introduced in Sec. \ref{sec:GAP}. In the simulations with the EPH model the size of the electronic subsystem is kept the same as the atomic subsystem and 4 grid points are included on x-, y- and z-directions. In all simulations, all types of possible outputs are fetched at an interval of 100 steps. The simualtion wall times of the TurboGAP code for the different simulations using different numbers of processors are shown in Table \ref{tab:TurboGAP-scaling}. %
\begin{table}[ht]
	\centering
	\caption{Scaling of the TurboGAP code with the implemented modules. M1 to M5 represent the "modes" of simulation, with the following description. M1: TurboGAP core without any module activated (reference); M2: only EPH model is activated; M3: only adaptive timestep is activated, M4: adaptive timestep and FES model are active; M5: adaptive timestep and EPH model are activated. For each mode, with the given simulation cell, 500 MD steps were performed. For M2 the timestep was 1 fs. For M3-M5, the maximum allowed timestep was 1 fs.}
	\label{tab:TurboGAP-scaling}
	\begin{ruledtabular}
		\renewcommand{\arraystretch}{1.2} 
		\begin{tabular}{l@{\hskip 8pt}c@{\hskip 8pt}c@{\hskip 8pt}c@{\hskip 8pt}c@{\hskip 8pt}c@{\hskip 8pt}c} 
			 & &\multicolumn{5}{c}{wall time (minutes)}\\ 
			 \cline{3-7}
			 Atoms & CPU & M1 & M2 & M3 & M4 & M5 \\
			 \hline
			$8\times10^{3}$ & 4 & 49 & 65 & 39 & 42 & 43 \\
			 & 8 & 33 & 33 & 23 & 21 & 22 \\
			 & 16 & 19 & 18 & 11 & 12 & 12 \\
			 & 24 & 11 & 12 & 7 & 7 & 8 \\
			\hline
			$10^{6}$ & 300 & 194 & 566 & 250 & 217& 577 \\
			 & 400 & 160 & 368 & 162 & 173 & 391 \\
			& 500 & 148 & 308 & 135 & 165 & 349 \\
			& 600 & 125 & 273 & 128 & 156 & 302 \\
	\end{tabular}
	\end{ruledtabular}
\end{table}
The simulations for 8000 atoms are performed using only one computing node, while for $10^6$ atoms cell, the processors are distributed over an arbitrary number of computing nodes in order to access the requested large number of processors. In general, the computation time for the same simulation using different nodes can vary to some extent and also, the performance of a specific computing unit varies depending on several real-time factors. As estimated, the times reported here can have a spread of about 20\%. The third column (M1) in Table \ref{tab:TurboGAP-scaling} is put as a reference, and shows the computation time when forces are calculated using the original TurboGAP core. As can be seen from each module, the performance scales up with increasing the number of processors. Therefore, the implemented modules maintain the original superior efficiency of TurboGAP for simulations with GAP MLIPs. 
 
\section{Bench-marking with LAMMPS}
\label{sec:Comparison of simulations using LAMMPS and TurboGAP}

As stated above, the TurboGAP code employs computationally more efficient ways to handle the SOAP and turboSOAP descriptors and therefore is able to achieve about an order of magnitude speedup in utilization of GAP MLIP models \cite{caro2019optimizing} compared to the foundational implementations of the descriptors in QUIP \cite{QUIP} or LAMMPS codes. The QUIP package can be used as a standalone tool to perform MD simulations. Additionally, the \verb|pair_style quip| command provides an interface for invoking QUIP routines within LAMMPS. Here we present benchmark comparisons between simulation results obtained from the new implementations in TurboGAP code and those obtained using LAMMPS.

\begin{figure*}[ht]
	\includegraphics[width = 0.9\textwidth]{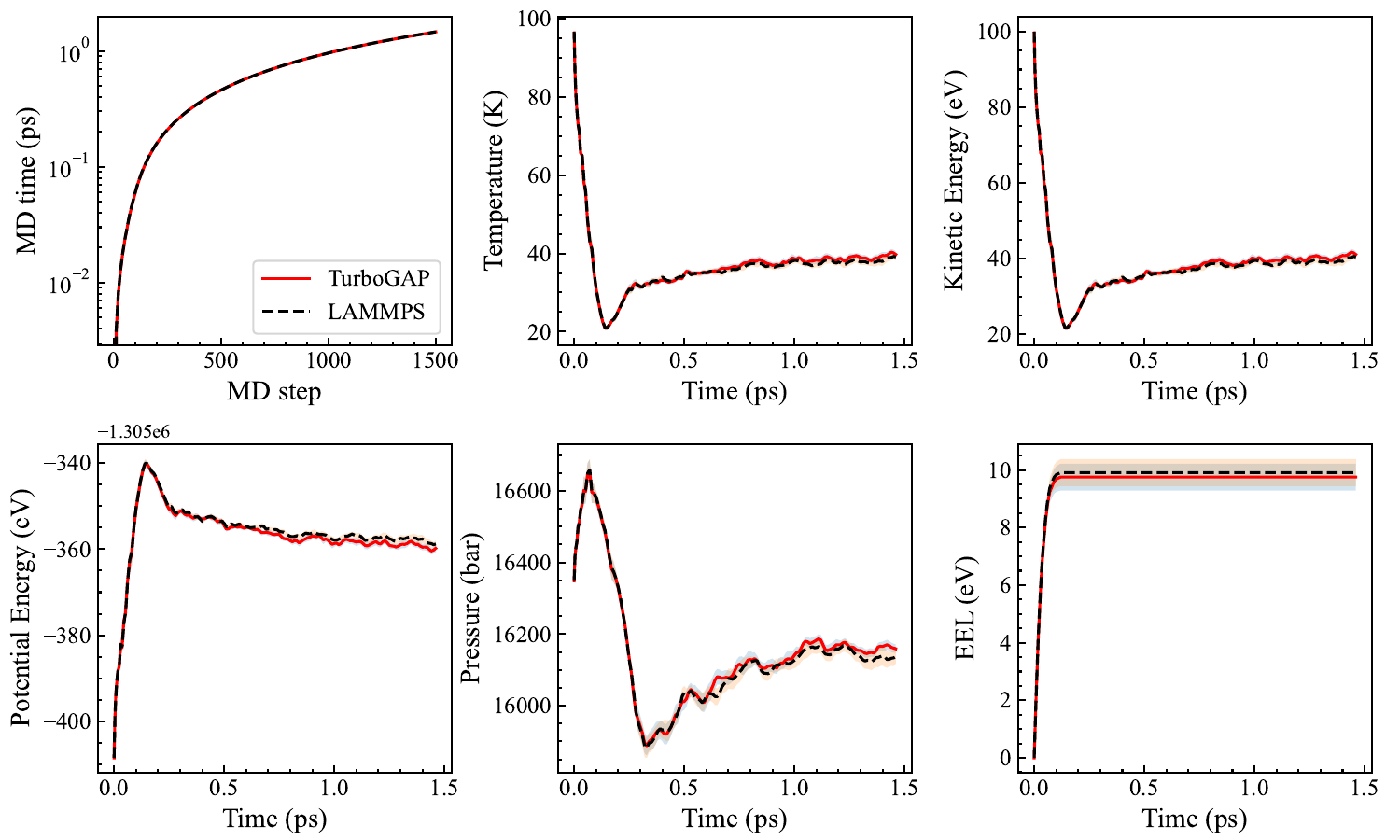}
	\caption{Comparison of the evolution of observables in 100 eV PKA cascade simulation with LAMMPS and implemented modules in TurboGAP. The results show the average over 10 random trajectories simulated with each code.}
	\label{fig:PKA-100eV-validation}
\end{figure*}

We compare the evolution of MD time, the thermodynamic properties, viz. temperature, pressure and energies, and the $E_{\mathrm{ES}}$ based on FES model in Si as a result of collision cascades initiated by a 100 eV PKA. Two identical simulations are initiated, where central atom in a system comprising of 4096 Si atoms at 0 K is given a velocity corresponding to 100 eV and the system is left to evolve for 1500 MD steps with adaptive calculation of timestep. For the adaptive timestep, we set $x_{\mathrm{max}}$, $e_{\mathrm{max}}$, $t_{\mathrm{max}}$, and $t_{\mathrm{min}}$ to be $0.1$ \AA,  $10$ eV,  $1$ fs and  $0.001$ fs, respectively. The cutoff energy for the FES model is 10 eV. The comparison of the outputs from two codes is presented in Fig. \ref{fig:PKA-100eV-validation}. 
A very good overall agreement of the time evolution of all quantities is observed. The cumulative $E_{\mathrm{ES}}$ obtained using TurboGAP is 15.314 eV and that using LAMMPS is 15.523 eV. This small difference has negligible effect when the statistical uncertainty coming from different random trajectories is taken into account (usually for a given PKA energy, several random trajectories are simulated and an average over all cases is reported for each property). Similar comparisons between the two codes with cutoff energy of 1 eV for $E_{\mathrm{ES}}$ and using multiple random trajectories are shown in the supplementary Fig. 2.

\begin{figure}[ht]
	\centering
	\includegraphics[width = 0.85\columnwidth]{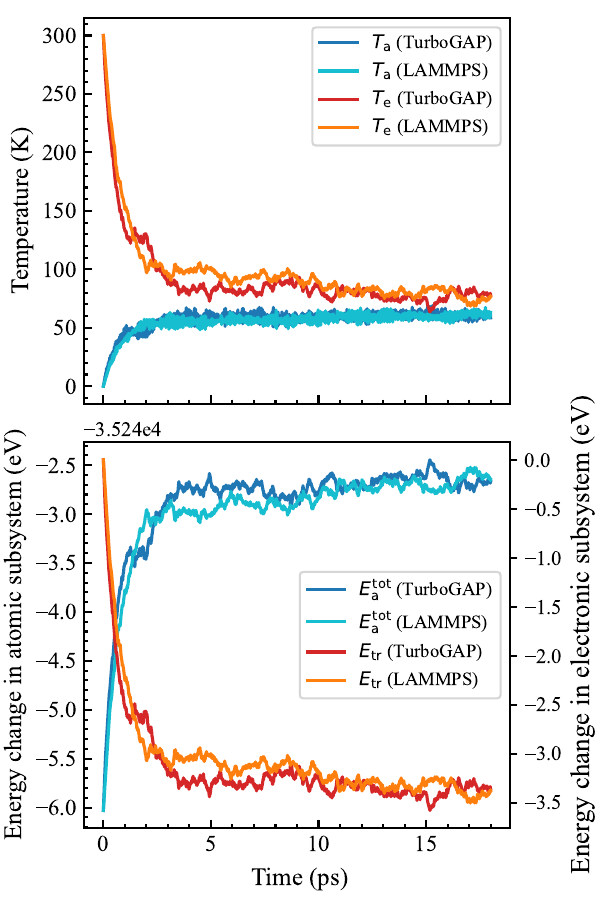}
	\caption{Comparison of the full EPH model simulations (inclusion of random and friction forces) using LAMMPS and TurboGAP codes. The electronic system is the same size as the atomic system which is 16.29 \AA \: on all three sides. The equilibration takes place at an intermediate temperature depending on the electronic heat capacity. Values of $C_e$ and $\kappa_e$ used are $3.5\times10^{-6} \; \mathrm{eV/K/\AA^3}$ and $0.1248 \; \mathrm{eV/K/\AA/ps}$, respectively.}
	\label{fig:fulleph-lammps-turbogap}
\end{figure}

Next, we compare the thermalization simulations using the EPH model in the two codes, see Fig. \ref{fig:fulleph-lammps-turbogap}.%
 In LAMMPS, the EPH model has been implemented as the "USER-EPH" plugin \cite{tamm2018langevin,tamm2019role}. Here again, the same input parameters and initial system configuration are maintained for simulations using both codes. The atomic system consists of 216 Si atoms with each side extending from 0 to 16.29 \AA \ and the electronic system is also of the same size. For all EPH simulations in this study, we used the constant type parametrization presented in Ref. \cite{jarrin2021integration}. It is seen that the temperature of the electron subsystem, $T_\mathrm{e}$, begins to fall rapidly towards the average equilibrium temperature whereas the temperature $T_\mathrm{a}$ of the atomic subsystem attains the equilibration slowly. On average, the temperatures of the electronic and atomic subsystems vary similarly in both TurboGAP and LAMMPS. Similar comparison was conducted between the two codes with the EPH model when only friction forces are active. Along with the GAP MLIP, the classical interatomic potentials were also considered in this comparison and the overall agreements are found to be good, see comparisons in supplementary Fig. 3.

\section{Simulations with EPH model and GAP MLIP}
\label{sec:Sim-Si-TurboGAP}

In this section, we present practical simulations using our implementations. First, we demonstrate the performance of the EPH model in equilibrating a test simulation cell, where it functions as a thermostat. We then investigate how the use of the FES and EPH models influences the simulation of primary damage. Finally, we illustrate how incorporating realistic ES from the EPH model, along with MLIPs, enhances the prediction of experimentally measured ion-beam mixing in silicon.

\subsection{Equilibration of atomic system using the EPH model}
\label{subsec:equil-with-eph-model}

\begin{figure}[htbp]
	\includegraphics[width = 0.9\columnwidth]{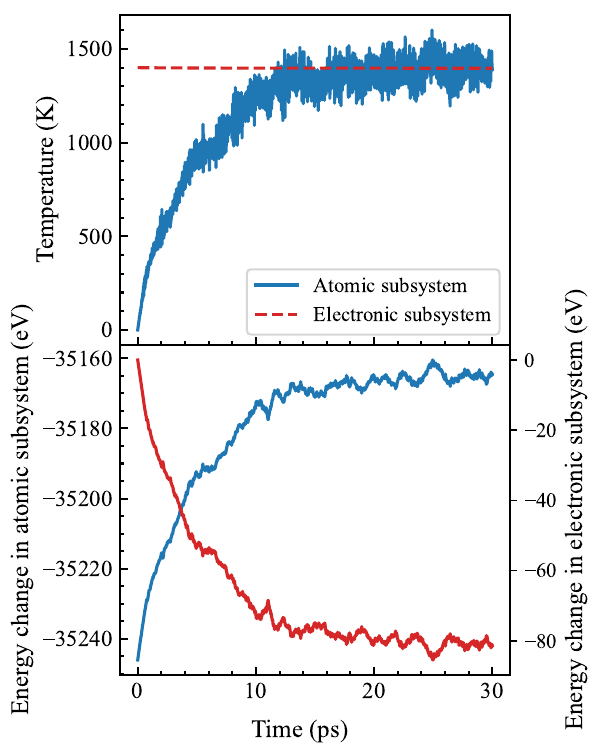}
	\caption{An illustration of equilibrating a system of 216 Si atoms with a sufficiently large electronic heat bath kept at 1400 K using the full EPH model as implemented in the TurboGAP code. With the constant ES coupling parameter the atomic system attains the equilibrium at 1400 K at about 15 ps. The evolution of the total energy of the atomic subsystem and the cumulative energy transfer from the electronic heat bath to the atoms show the conservation of energy in the combined atom-electron system.}
	\label{fig:full-eph-turbogap}
\end{figure}

Solving the heat diffusion equation, Eq. (\ref{eqn:hdeqn}), for the electronic subsystem controls $T_\mathrm{e}$, which in turn, will have effect on the random forces. To model the electronic system as an efficient heat reservoir or thermostat (further discussion in Sec. \ref{subsubsec:sim-setup}), its size should be larger than the atomic subsystem. An illustration of applying the complete EPH model to achieve thermal equilibrium of the atoms with a very large electronic heat bath using the TurboGAP code is shown in Fig. \ref{fig:full-eph-turbogap}. %

Here, the atomic subsystem with 216 Si-atoms is initially at 0 K and the temperature of the electronic subsystem is initially kept at 1400 K and the total system is evolved for 30 ps. The atomic temperature, $T_\mathrm{a}$, attains an average value of 1400 K after about 15-18 ps. This rate of reaching equilibrium depends on the coupling parameter. The evolution of energies in the atomic and electronic subsystems are shown on the bottom panel of Fig. \ref{fig:full-eph-turbogap}. As the atomic subsystem comes to equilibrium around 1400 K, its total energy changes by approximately 80 eV. This energy is supplied by the electronic subsystem which is seen from the cumulative energy transfer between the two subsystems shown on the right hand y-axis of the plot. To assess the sensitivity of the implementation of the EPH model and the efficiency or the rate of equilibration at different temperatures, a similar analysis with lower temperatures for the electronic subsystem was performed. The results of this analysis are presented in Supplementary Fig. 4.

It is important to note that the specific heat capacity and conductivity of the electronic subsystem in silicon are small \cite{glassbrenner1964thermal,koroleva2017determination}, and using such small parameters introduces instability in the solution of the heat diffusion equation. Moreover, such low values for these parameters are not efficient enough to maintain a constant temperature in the electronic subsystem which is required in case of a heat reservoir. For the purposes of the present study, we intend to model the electronic subsystem primarily as a heat reservoir while simulating radiation-induced cascades in the atomic subsystem (see \ref{subsec:R-ICC}) and for that, we use sufficiently high values of the electronic parameters, not close to the real ones.

It is noteworthy here that activating only friction in the EPH model is not equivalent to the calculation of $E_{\mathrm{ES}}$ using the FES model. The tensor form of the electronic friction in EPH model acts on the relative velocities of the atoms and dampens their motion in a different way than in the FES model. On a separate note, even with the EPH model, however, one can reflect upon the fact that activating only the friction forces without the random forces would deviate from the fluctuation-dissipation principle, which states that the friction to an external disturbance results from the intrinsic random fluctuations in the system \cite{kubo1966fluctuation}. The EPH model uses only one free parameter which is the coupling strength as a function of electron density. The electron densities are used as weights to the relative random and friction forces on the atoms. The model has been developed for metals and it is most suitable to study the processes far from equilibrium as is the case with radiation cascades in materials. In the MD study of collision cascades, this model provides excellent agreement with the first-principle methods. It considers that the ES is linear with velocity. For very low energies, this seems to hold quite satisfactorily in case of metals, but can be different in case of semiconductors \cite{lim2016electron}. Such low energies are however of no interest for radiation-induced cascade simulations performed in this study.

\subsection{Cascade simulations and Ion-beam mixing in Si}
\label{subsec:R-ICC}

In this section we present cascade simulations in Si, using the GAP MLIP and the EPH model. From our simulations, we calculate the ion-beam mixing values and compare it to experimental values reported in the literature. We also preform the same set of simulations with the FES model. This allows to investigate how the incorporation of the EPH model affects the description of the electronic effects and hence the evolution of the irradiation-induced defects. All simulations are conducted with the TurboGAP code, using our implemented modules. The highest PKA energy simulated is 10 keV which is launched in a $10^{6}$ atoms cell. First, the details of the simulation setup are presented. Then, the difference in predictions of the FES and EPH models in cascade simulations is investigated. Finally, the results of the mixing values are reported and discussed.

\subsubsection{Simulation setup}
\label{subsubsec:sim-setup}

The sizes of the simulation cells and the number of random directions for different PKA energies are provided in Table \ref{tab:Sim-setup-results}. 
\begin{table}[ht]
	\centering
	\caption{Details of the cascade simulations. Size of the system (number of atoms) was selected based on the PKA energy. All the simulation cells are cubic. The last column represents the number of independent random directions simulated to account for statistical variability in the results.}
	\label{tab:Sim-setup-results}
	\begin{ruledtabular}
		\renewcommand{\arraystretch}{1.3} 
		\begin{tabular}{c@{\hskip 8pt}c@{\hskip 8pt}c@{\hskip 8pt}c} 
			PKA (keV) & atoms & ES  & cases \\
			\hline
			 0.1 & 4096 & EPH, FES & 20, 20\textsuperscript{a} \\
			0.2 & 8000 & EPH & 20 \\
			0.4 & 17576 & EPH & 20 \\
			1.0 & 46656 & EPH & 20 \\
			2.0 & 85184 & EPH, FES & 4$\times$100\textsuperscript{b}, 2$\times$100\textsuperscript{c}\\
			5.0 & 512000 & EPH & 20 \\
			10.0 & 1000000 & EPH, FES & 20, 2$\times$20\textsuperscript{d}
		\end{tabular}
	\footnotetext{Kinetic energy cutoff of 1 eV in FES model}
	\footnotetext{100 cases for each one of the EPH model settings. Four settings were tested (see supplementary material).}
	\footnotetext{100 cases for each one of the FES cutoff values, 1 and 10 eV.}
	\footnotetext{20 cases for each one of the FES cutoff values, 1 and 10 eV.}
	\end{ruledtabular}
\end{table}
For each energy, the pristine cell was equilibrated at 300 K and zero pressure using the Berendsen thermostat and barostat \cite{berendsen1984molecular,bussi2007canonical}. All cells were equilibrated for at least 20 ps. The thermostat functioning of the electronic subsystem in the EPH model can be achieved with its size maintained at approximately four times that of the atomic subsystem. Additionally, the temperature gradient within the electronic subsystem can be accounted for by imposing a grid within the electronic subsystem (Eq. \ref{eqn:hdeqn}). In simulations incorporating a grid, the electronic subsystem is divided in three dimensions, with a mesh size of 20–22 \AA. The initial temperature of the electronic subsystem is set to 300 K. The full EPH model, i.e including both random and friction forces, is used throughout to simulate the cascades. The values of $C_\mathrm{e}$ and $\kappa_\mathrm{e}$ may be either constant or temperature-dependent. For constant $C_\mathrm{e}$, a value of $9.738 \times 10^{-5} \, \mathrm{eV/K/\AA^3}$ was used. For the constant case of $\kappa_\mathrm{e}$, it was set to $9.738 \times 10^{-2} \, \mathrm{eV/K/\AA/ps}$, based on Ref. \cite{glassbrenner1964thermal}. As discussed in Sec. \ref{subsec:equil-with-eph-model}, these values are very large compared to the actual electronic parameters in order for the electronic subsystem to function as an effective thermostat. For the FES model, data from SRIM-2013 was used and a border cooling layer with a thickness of 2 \AA \: was applied in all directions, maintaining a constant temperature of 300 K using the Berendsen thermostat. For the adaptive timestep algorithm, $x_{\mathrm{max}}$ and $e_{\mathrm{max}}$ were set to 0.1 \AA and 30 eV, respectively, with a maximum allowed timestep of 1 fs. The cascade simulations were carried out for 10 ps, at which point the number of surviving defects saturated. The uncertainty in any quantity reported here is the standard error. 

\subsubsection{Comparing dissipation models}
\label{subsubsec:eph-settings}
In this section, we investigate how the choice of dissipation model affects the prediction of primary damage. For this purpose, a 2 keV PKA is selected, and the predictions of the FES and EPH models are compared. With the FES model, simulations were performed with both 1 eV and 10 eV cutoff for the kinetic energy. In the simulations with the EPH model, the electronic subsystem functions as a thermostat to maintain the atomic subsystem at 300 K, the temperature gradient within the electronic subsystem is taken into account, and $C_\mathrm{e}$ and $\kappa_\mathrm{e}$ vary as a function of temperature. To capture statistical variability in the results, 100 random PKA directions for each setup were simulated and the average values for the observables are reported. The EPH model is applicable with different setups. Supplementary Fig. 5 and Supplementary Fig. 6 illustrate the effect of these setups on the predicted primary damage. 

\begin{figure*}[ht]
	\includegraphics[width = 1.0\textwidth]{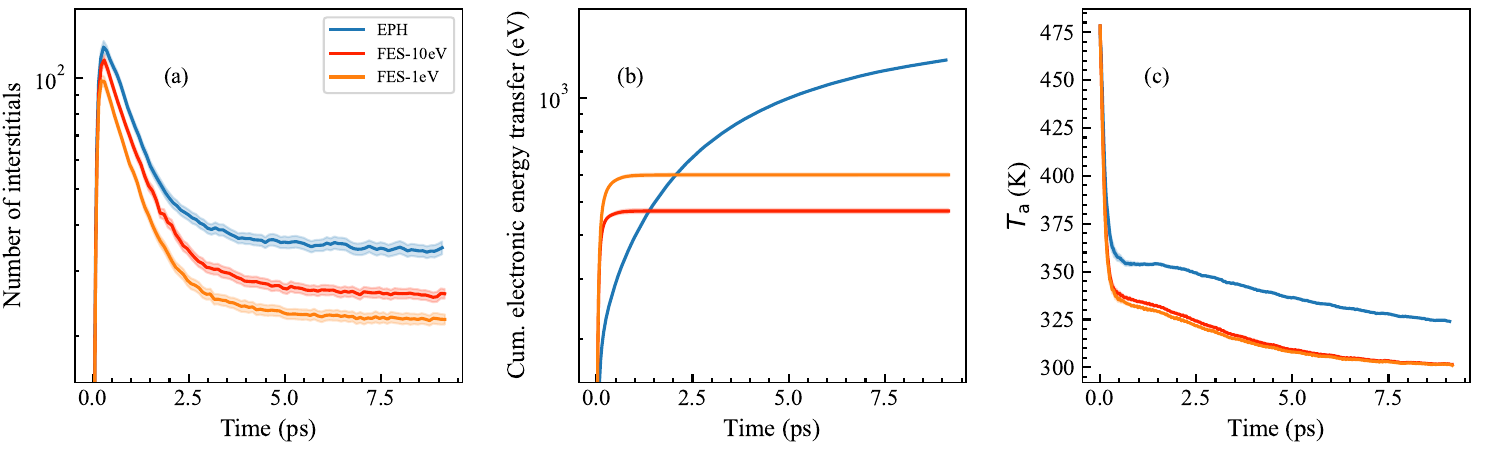}
	\caption{Time evolution of the (a) number of interstitials; (b) cumulative energy transferred to the electronic system, and (c) temperature in the atomic subsystem due to 2.0 keV Si self-ion cascades. The shaded area on each graph represents the standard error of the mean values in the plot.}
	\label{fig:pointdefects-plot-only-2keV}
\end{figure*}

The plots in Fig. \ref{fig:pointdefects-plot-only-2keV} illustrate the time evolution of key quantities related to the atomic and electronic subsystems during
 cascades induced by a 2 keV PKA simulation. As shown in Fig. \ref{fig:pointdefects-plot-only-2keV}a, the number of interstitials at the heat spike is higher in the EPH model compared to the FES model. Among the FES models (with different cutoff values), the model with the higher cutoff consistently produces a greater number of defects. Furthermore, the number of surviving defects is higher in the EPH model than in the FES models. On average, the EPH model yields 33 surviving defects, whereas the FES models with cutoff values of 10.0 eV and 1.0 eV result in 26 and 22 surviving defects, respectively.

The cumulative energy transferred to the electronic system differs considerably between the two ES models, as shown in Fig. \ref{fig:pointdefects-plot-only-2keV}b. In both models, the energy initially increases at nearly the same rate. While the accumulated energy saturates in the FES model, it continues to increase gradually in the EPH model. In the FES model, atoms with kinetic energies above the cutoff experience retarding forces, and as the cascade cools, the kinetic energies of all atoms eventually drop below this threshold. This leads to no further increase in the accumulated energy due to the dissipation by the applied friction forces on the atoms. The same logic can be used to explain the lower defect count in the FES-1eV model compared to the FES-10eV model. The kinetic energy cutoff can impact the re-crystallization rate of the atoms in the disordered core of the cascade. A lower kinetic energy cutoff results in smaller displacements of the energetic atoms involved in cascades, which in turn reduces defect creation.

\begin{figure}[ht]
	\includegraphics[width = 1.0\columnwidth]{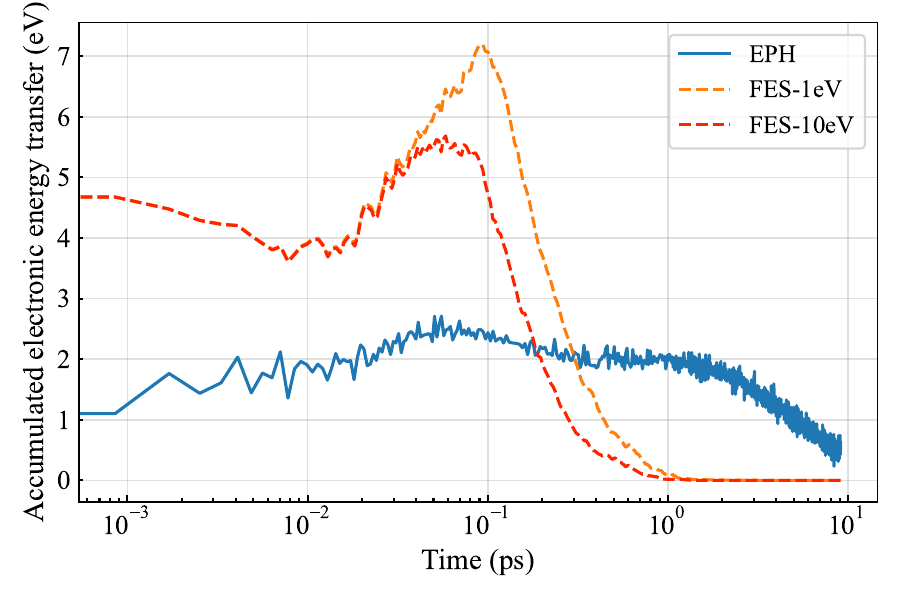}
	\caption{Energy transfer to the electronic system accumulated over every 10 MD steps. The results are from 2 keV PKA cascade simulations while using FES and EPH models. The plots present average values over 100 cases in each model.}
	\label{fig:2keV-energy-dissipations-eph-SRIM}
\end{figure}

For both the FES and EPH models, we calculated the accumulated energy transfer to the electronic system over the fixed intervals of 10 MD steps, shown in Fig. \ref{fig:2keV-energy-dissipations-eph-SRIM}. In the FES model, the energy dissipation peaks and then decays to zero as the cascade cools down, which 
results in a saturated total accumulated energy profile in Fig. \ref{fig:pointdefects-plot-only-2keV}b. The EPH model, in contrast, continues to capture energy transfer to the electronic system in the post-cascade regime through e-ph coupling. This behavior of the EPH model, illustrated in Fig. \ref{fig:2keV-energy-dissipations-eph-SRIM}, explains the non-saturated total cumulative energy seen in Fig. \ref{fig:pointdefects-plot-only-2keV}b. Supplementary Fig. 6 illustrates the same quantity shown in Fig. \ref{fig:2keV-energy-dissipations-eph-SRIM} for different setups of the EPH model. This figure highlights a key advantage of the EPH model; its independence from an arbitrarily selected cutoff value, a limitation of the FES model. While the transferred energy in the FES model is sensitive to the kinetic energy cutoff, Supplementary Fig. 6 shows that different settings of the EPH model predict very similar energy transfer. Supplementary Fig. 5 and Supplementary Fig. 6 show that the evolution of defects, energies and temperatures is mostly independent of small variations in the setup of electronic subsystem used as a thermostat.

 The evolution of temperature in the atomic subsystem is shown in Fig. \ref{fig:pointdefects-plot-only-2keV}c. In case of the FES model, the use of Berendsen thermostat is more efficient than thermostating with EPH to cool down the atomic subsystem to the desired temperature. With the EPH model, the temperature of the atomic subsystem reaches around 330 K after 7 ps. In other words, it takes longer for the atomic system to cool down to 300 K, as observed in Sec. \ref{subsec:equil-with-eph-model}. It is to be noted that, even though the atomic temperature and the cumulative $E_{\mathrm{ES}}$ after heat spike is higher with the EPH model, the number of surviving point defects is not lower compared to the case when the FES model (either with 1 eV or 10 eV cutoff) is used. 

 To investigate whether this difference in equilibrium influences the number of surviving defects, we re-ran our cascade simulations for the FES-1eV and FES-10eV models, setting the border temperature to 324 K. Supplementary Fig. 5a confirms that these revised FES-1eV and FES-10eV simulations also equilibrated at 324 K, matching the EPH model’s temperature. Additionally, Supplementary Fig. 5b shows that the 24 K shift in equilibrium temperature does not alter the number of surviving defects in the FES-1eV and FES-10eV models. This shows that there is not much effect of recombination of defects due to diffusion at higher temperature ($\sim$ 24 K higher), and the number of surviving defects predominantly carries the signature of numbers of defects formed during the heat spike phase. The higher ratio of number of defects in EPH compared to FES model remains same throughout the simulation.

Apart from the number of generated defects, we investigated the effect of predicted ESP from the EPH and FES models on the clustering of the defects. The clustering of defects caused by radiation and the distribution of their sizes play a key role in shaping the microstructural changes in materials under irradiation. The dimensions and spacing of these clusters influence their longevity and the intensity of their elastic interactions. Consequently, understanding the size distribution of clusters is essential for models like cluster dynamics or rate theory \cite{rate2,rate1}. Also, in nanoelectronics, the interaction between low-energy implanted dopants and defect clusters poses a challenge, contributing to diffusion and activation anomalies that hinder achieving the desired component design. A precise model of the surviving defects’ evolution and structure could enhance our understanding of dopant-defect interactions, enabling a more accurate approach to addressing this phenomenon \cite{bazizi2010transfer,martin2006modeling,koelling2013direct}. The cluster analysis was performed with the dedicated module in the OVITO \cite{ovito} software. A defect cluster consists of a group of adjacent Wigner–Seitz defects, where neighbors are defined as defects situated within a cutoff distance of 5.3 \AA \cite{nordlund1998defect}, nearly the lattice constant of Si. We considered both the surviving defects and the defects generated at the heat spike phase. 

\begin{figure}[ht]
	\includegraphics[width = 0.9\columnwidth]{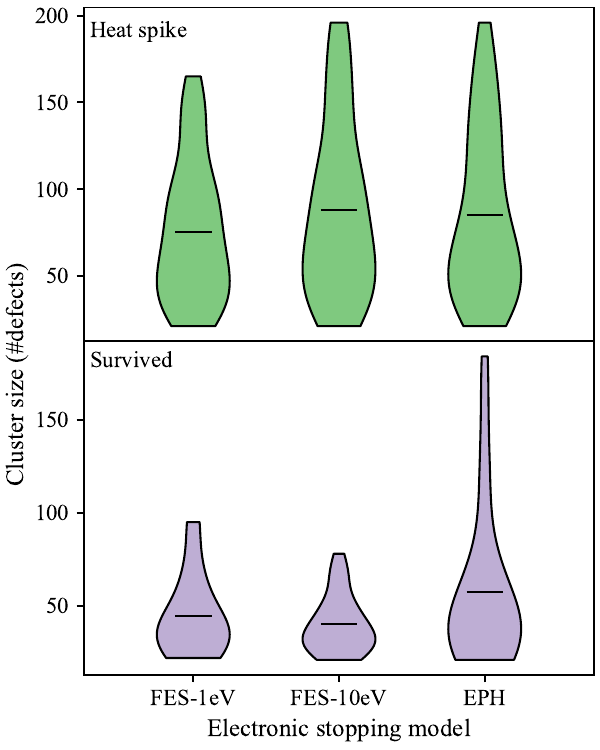}
	\caption{Size distribution of the defect clusters in 2 keV PKA cascade simulations with different ES models. Only clusters with sizes exceeding 20 interstitials have been considered in the distribution analysis. The horizontal line within each violin shows the mean value of the distribution.}
	\label{fig:clustering}
\end{figure}

Figure \ref{fig:clustering} compares the size distribution of defect clusters derived from our cascade simulations using the FES-1eV, FES-10eV, and EPH models.
 Here, 'cluster size' refers to the number of interstitials within each cluster. The distribution in Fig. \ref{fig:clustering} includes only clusters with sizes exceeding 20 interstitials. While the pocketing of defects during the heat spike remains largely unaffected by the ESP models, the clustering of surviving defects exhibits significant variation. As seen in Fig. \ref{fig:clustering}, the average size of clusters in the EPH model is larger than that in FES models. This is consistent with the observed size distributions: while the largest cluster sizes in FES models remain below 100, EPH model contains clusters exceeding this size, thereby increasing the overall average.

To determine if the difference in the equilibrium temperature of cascades can affect the configuration of defect clusters, we re-calculated the size distribution of clusters from revised FES-1eV and FES-10eV cascade simulations at 324 K. The size distribution of defect clusters from the FES-based cascades at 324 K is presented in Supplementary Fig. 8 alongside the EPH model’s results. Both Figure \ref{fig:clustering} and Supplementary Fig. 6 reveal consistent clustering trends between the EPH and FES models, indicating that equilibrium temperature has no significant impact on the clustering of surviving defects. Thus, the observed differences arise primarily from the models themselves.

\subsubsection{Atomic mixing}
\label{subsubsec:at-mix}
Ion-beam mixing is one of the few observables that can be calculated from atomistic radiation damage simulations, and directly compared against experimentally measured values. When cascade overlapping is insignificant, the average mixing, which is determined by weighting the mixing caused by primary recoils and integrating over the incident ion's recoil spectrum, can be directly compared to the mixing measured from experiments \cite{andrea_nickel, deb3}. Accordingly, we compare the mixing values obtained from our simulations with the experimental mixing values to evaluate our simulations and results. The atomic mixing is calculated according to Eq. \ref{eqn:mixingQ} \cite{nordlund1998mechanisms} %
 \begin{subequations}
 	\label{eqn:mixingQ}
 	\begin{equation}
 		\label{eqn:Q-from-Rsquare}
 		Q = \frac{R^2}{6n_0E_D},
 	\end{equation}
 	\begin{equation}
 		\label{eqn:R-square-in-Q}
 		R^2 = \sum_i [\mathbf{r}_i(t) - \mathbf{r}_i(t = 0)]^2
 	\end{equation}
 \end{subequations}
 where $R^2$ is the square of the atomic displacements; $n_0$ is the atomic density, and $E_D$ is the damage energy or deposited nuclear energy. The quantity $R^2$ is calculated as a function of time according to the Eq. \ref{eqn:R-square-in-Q} where $\mathbf{r}_i$ is the positional vector of atom $i$ at the time $t$. To calculate $R^2$, we used the EPH model (the same settings as presented in Sec. \ref{subsubsec:eph-settings}) to simulate the primary damage in Si with the PKA energies in the range of 0.1-10 keV. The maximum number of defects during the heat spike and the number of surviving defects obtained from these simulations are provided in Table \ref{tab:point-defect-statistics}. %
 \begin{table}[ht]
 	\centering
 	\caption{Defect states in a self-ion irradiated Si, obtained in our cascade simulations with different PKA energies and two ES models, EPH and FES. $t_{\mathrm{spike}}$ is the average time at which the heat spike develops. $N_{\mathrm{max}}^{\mathrm{int}}$ is the number of interstitials at the heat spike. $N_{\mathrm{stable}}^{\mathrm{int}}$ is the number of surviving defects.}
 	\label{tab:point-defect-statistics}
 	\begin{ruledtabular}
 		\renewcommand{\arraystretch}{1.3} 
 		\begin{tabular}{l@{\hskip 8pt}c@{\hskip 8pt}c@{\hskip 8pt}c@{\hskip 8pt}c} 
 			ES  & PKA(keV) & $t_{\mathrm{spike}}$ & $N_{\mathrm{max}}^{\mathrm{int}}$ & $N_{\mathrm{stable}}^{\mathrm{int}}$  \\
 			\hline
 			EPH & 0.1 & $0.16\pm0.002$ & $9\pm0.3$ & $2\pm0.2$ \\
 			& 0.2 & $0.21\pm0.005$ & $17\pm0.6$ & $2\pm0.2$ \\
 			& 0.4 & $0.2\pm0.008$ & $35\pm1.2$ & $5\pm0.3$  \\
 			& 1.0 & $0.24\pm0.022$ & $64\pm4.8$ & $15\pm1$  \\
 			& 2.0 & $0.29\pm0.014$ & $122\pm4.7$ & $35\pm1.4$ \\
 			& 5.0 & $0.36\pm0.051$ & $280\pm21$ & $83\pm4$ \\
 			& 10.0 & $0.69\pm0.129$ & $508\pm39$ & $182\pm8$ \\
 			\hline
 			FES-1eV & 2.0 & $0.23\pm0.011$ & $98\pm3.7$ & $22\pm0.7$ \\
 			& 10.0 & $0.74\pm0.13$ & $327\pm23$ & $104\pm4$  \\
 			\hline			    
 			FES-10eV & 0.1 & $0.16\pm0.002$ & $8\pm0.4$ & $1\pm0.2$ \\
 			& 2.0 & $0.3\pm0.013$ & $112\pm3.4$ & $26\pm0.8$ \\
 			& 10.0 & $0.5\pm0.091$ & $390\pm38$ & $136\pm7$ \\		
 		\end{tabular}
 	\end{ruledtabular}
 \end{table}

\begin{figure}[ht]
	\includegraphics[width = 1.0\columnwidth]{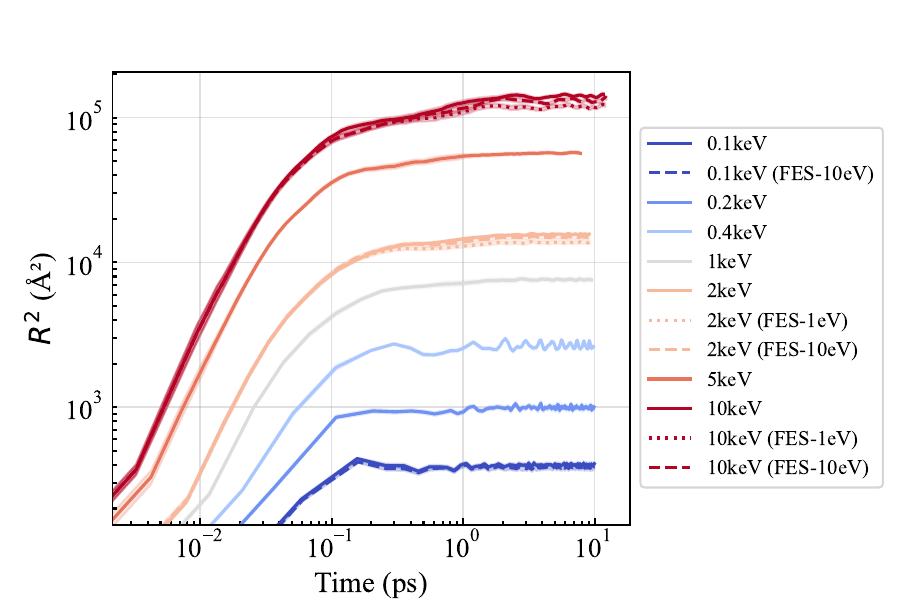}
	\caption{The quantity $R^2$ (Eq. \ref{eqn:R-square-in-Q}) as a function of time in the collision cascades induced by 0.1-10 keV Si PKAs. The mean values and uncertainties shown are calculated from 20 (100 cases for 2 keV PKA) sets of cascade events simulated using the TurboGAP code.}
	\label{fig:R-square-with-times-all-energies}
\end{figure}

 For a particular PKA energy, Q is computed from the average value of $R^2$ over full time of cascade evolution (8 - 10 ps). The average variations of $R^2$ for different energies from 0.1 to 10 keV (averaged over different random directions of PKA trajectories) with time are shown in Fig. \ref{fig:R-square-with-times-all-energies}. 
 The variation of mixing is governed by the magnitude of atomic displacements. The displacements of the atoms 
increase rapidly immediately after the onset of a cascade event and flatten out by about 1.5 ps. For PKAs with a few hundreds of eV, 80-90\% of mixing is complete by this time, while for high energy PKAs (5-10 keV) it is only 60-70\% complete. Using the EPH model, the atomic displacements are slightly larger than in case of using the FES model mainly due to the lower $E_{\mathrm{ES}}$ experienced by the ions around the heat spike phase in EPH model (see Fig. \ref{fig:2keV-energy-dissipations-eph-SRIM}). Moreover, the difference in $R^2$ between the ES models becomes greater as the PKA energy increases. In addition, the use of different cutoff values for the FES model leads to different amounts of displacements of the atoms, where, as expected, setting a higher cutoff value produces larger mixing.

The damage energies ($E_D$), with which the mixing values were calculated, are compiled in Table \ref{tab:edn}. The $E_\mathrm{D}$ was calculated with SRIM-2013 
  \begin{table}[ht]
 	\centering
 	\caption{The damage energy (self irradiated) for different Si PKA energies used for the calculation of atomic mixing from our cascade simulations.}
 	\label{tab:edn}
 	\begin{ruledtabular}
 		\renewcommand{\arraystretch}{1.3} 
 		\begin{tabular}{cc}
 			PKA (keV)  & $E_\mathrm{D}$ (eV)\\
 			\hline
 			0.1 & 83 \\
 			0.2 & 162 \\
 			0.4 & 314 \\
 			1.0 & 754 \\
 			2.0 & 1455 \\
 			5.0 & 3433 \\
 			10.0 & 6476
 		\end{tabular}
 	\end{ruledtabular}
 \end{table}
using the method introduced in Ref. \cite{stoller2013use}. The damage energy is defined as $E_\mathrm{D}=E_\mathrm{PKA}-E_\mathrm{ionize}$, where $E_\mathrm{ionize}$ is the energy lost to ionization of target (Si) atoms. For each PKA energy, SRIM was run for self-irradiation of Si using the "Quick" Kinchin and Pease option. The recommended threshold displacement energy was used and the binding energy was set to zero. The $E_\mathrm{ionize}$ was calculated by integrating the $E_{\mathrm{ES}}$ over target layer thickness. The $E_{\mathrm{ES}}$ is strored in SRIM's "IONIZ.txt" output file. The integration was done for both the incident and recoil atoms, and summed to obtain the $E_\mathrm{ionize}$.
 
 The calculated mixing values are compared to the values reported in two other simulation works \cite{nordlund1998mechanisms,hamedani2021primary} in Table \ref{tab:mixingQ-self-ion-simulations}. In Ref. \cite{hamedani2021primary} the 
 \begin{table}[ht]
 	\centering
 	\caption{Ion-beam mixing from cascade simulations in this work, compared to the simulations in the literature.} 
 	\label{tab:mixingQ-self-ion-simulations}
 	\begin{ruledtabular}
 		\renewcommand{\arraystretch}{1.3} 
 		\begin{tabular}{lccc} 
 			&  & \multicolumn{2}{c}{Q ($\AA^5$/eV)} \\ 
 			  \cline{3-4}
 			   ES & PKA(keV)   & this work & literature  \\
 			\hline
 			EPH 		&   0.1 & $16.3 \pm 0.1$  	 &		$18 \pm 1$\textsuperscript{a} \\
                            &	0.2 & $20.7 \pm 0.2$	&	  $21 \pm 1$ \textsuperscript{a}\\
                            &	0.4 & $28.3 \pm 0.4$ 	&		$24 \pm 1$\textsuperscript{a}, $6 \pm 11$\textsuperscript{b} \\
                            &	1.0 & $32.8 \pm 0.6$	&		$32 \pm 1$\textsuperscript{a} \\
                            &	2.0  & $33.8 \pm 0.8$, 	&		$36 \pm 1$\textsuperscript{a}, $16 \pm 2$\textsuperscript{b} \\
                            &	5.0  & $48.3 \pm 1.6$	&		\\
                            &	10.0  & $60.7 \pm 1.8$	&		$15 \pm 3$\textsuperscript{b} \\
 			\hline
 			FES-1eV 	&  2.0  &  $30 \pm 0.7$  & \\
				 			& 10.0 &  $52.9 \pm 1.5$ & \\
 			\hline
 			FES-10eV   &  0.1  &  $15.7 \pm 0.1$  & \\
				 			&  2.0  &  $32.4 \pm 0.7$  & \\
 							&  10.0  &  $57.6 \pm 1.7$
 		\end{tabular}
 		\footnotetext{Ref. \cite{hamedani2021primary} (mod-GAP)}
 		\footnotetext{Ref. \cite{nordlund1998mechanisms}}
 	\end{ruledtabular}
 \end{table}
 DMol repulsive potential \cite{nordlund1997repulsive} and FES-10eV model have been used in the cascade simulations. Our mixing values reported in Table \ref{tab:mixingQ-self-ion-simulations} are estimated using all our simulation cases (Table \ref{tab:Sim-setup-results}). The simulations in \cite{nordlund1998mechanisms} were carried out using Stillinger-Weber (SW) potential. 

Table \ref{tab:mixingQ-self-ion-simulations} shows that there is not a considerable difference in the total atomic mixing between the EPH and FES models. This is expected, as a similar trend is observed for atomic displacements (Fig. \ref{fig:R-square-with-times-all-energies}). When compared with Ref. \cite{hamedani2021primary}, a good agreement is observed with the outcome of present simulations using the FES model. It is to be noted that, in the present work, the mixing values reported are those obtained by averaging the mixing values over time whereas the values from Ref. \cite{hamedani2021primary} are those at $t\sim$2 ps. Predictions from simulations using the GAP MLIP differ significantly (by a factor of 2 to 4) compared to those obtained using the SW potential in Ref. \cite{nordlund1998mechanisms}. 
 
We compare the atomic mixing obtained from our simulations with experimentally measured data from \cite{matteson1981ion} and \cite{paine1981comparison} as well. These experimental data are from irradiation of Si with Ne, Kr, Ar, and Xe ions in the energy range of 50-300 keV and with different marker elements. In an irradiation event there is a spectrum of energy of the primary recoils formed due to the incident ion in the target. For the purpose of comparison, the fractional distribution of the recoil energies should also be taken into account. Hence, in the calculation of the mixing values, a weighted sum over the spectrum is performed which is expressed in 
Eq. \ref{eqn:mixingQ-weighted}:

\begin{equation}
	\label{eqn:mixingQ-weighted}
	Q = \frac{\sum_{E} R^2(E)N(E)\Delta{E}}{6n_0E_D}
\end{equation}

where $R^2(E)$ is obtained from Eq. \ref{eqn:R-square-in-Q} and then averaged over time; $N(E)$ is the fraction of recoils; $\Delta{E}$ is the width of an energy bin in the spectrum, and $E_D$ is the deposited nuclear energy in Si calculated for the incident ion (type and anergy) used in experiment.

\begin{figure}[ht]
	\includegraphics[width = 1.0\columnwidth]{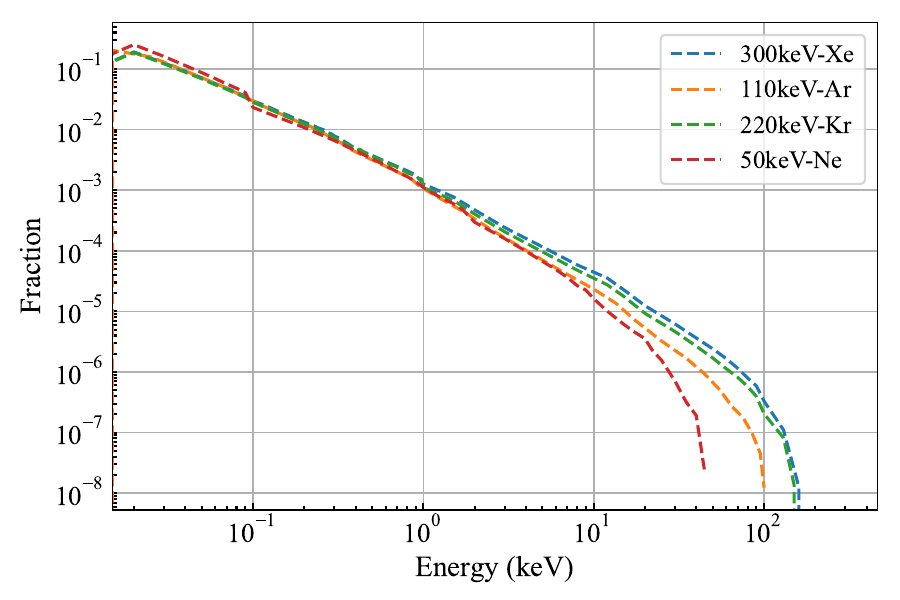}
	\caption{PKA spectra (normalized to unity) for irradiation of Si with Xe, Ar, Kr, and Ne. The data is acquired from SRIM-2013.}
	\label{fig:PKA-spectra-experiments}
\end{figure}

The primary recoil spectra for different incident ions, calculated using the SRIM-2013 code \cite{uasaha2023} is shown in Fig. \ref{fig:PKA-spectra-experiments}. %
Since our MD simulations cover only a limited selection of Si PKA energies, we bin the spectrum based on the simulated PKA energies (0.1-10 keV) where the number of bins matches the number of PKA energies. The last energy bin is made large enough to contain all recoils with energies greater than or equal to 8 keV. This somewhat compensates the lack of high energy PKAs greater than 10 keV in our simulation. From Fig. \ref{fig:PKA-spectra-experiments} it is seen that the resulting PKA spectrum for any of the combinations of incident ion and energy is similar and has a decreasing fraction of recoils at higher energies. 

The comparison of the calculated mixing values from simulations with the experimental counterparts are given in Table \ref{tab:mixingQ-allE-experiments}.  \begin{table}[ht]
	\centering
	\caption{Comparison of atomic mixing values from simulations using the EPH model and from implantation experiments. The target material in experiments is silicon. $E_\mathrm{D}$ is the damage energy.}
	\label{tab:mixingQ-allE-experiments}
	\begin{ruledtabular}
		\renewcommand{\arraystretch}{1.3} 
		\begin{tabular}{lccc}
			Ion, energy(keV)  & $E_\mathrm{D}$(keV)  & Q simulation     &  Q experiment \\
									 & 							 		& 	($\AA^5$/eV)  &  ($\AA^5$/eV)  \\
			\hline
			Ar, 110  				& 49.786 						& $62 \pm 3$  &  $58 \pm 32$ \footnotemark[1] \\
			Kr, 220  				& 114.796 					   & $40 \pm 1$  & $34 \pm 6$ \footnotemark[2] \\
			Xe, 300  				& 160.56   					   &  $34 \pm 1$  & $44 \pm 7$ \footnotemark[2] \\
			Ne, 50   				& 23.155   						&  $80 \pm 5$  & $40 \pm 13$ \footnotemark[2] \\
		\end{tabular}
            \footnotetext[1]{Ref. \cite{paine1981comparison}}
		\footnotetext[2]{Ref. \cite{matteson1981ion}}
	\end{ruledtabular}
\end{table}
The experimental data shown here are taken from Refs. \cite{paine1981comparison,matteson1981ion} which provide the mixing obtained in Si using Ge as marker element. In the experiments, mixing is quantified by measuring the broadening of the mixing layer following ion irradiation.

The mixing values from the EPH model are in good agreement with the experiments. Regarding the accuracy of the results from simulations, the effect of the ES model and the interatomic potential could be highlighted. We have not performed cascade simulations for all the selected PKA energies with the FES model (Table \ref{tab:point-defect-statistics}). However, the FES model is expected to yield mixing values comparable to those acquired with the EPH model, reported in Table \ref{tab:mixingQ-allE-experiments}. This expectation is supported by taking into account the Q values for the representative 2 and 10 keV PKA energies in Table \ref{tab:mixingQ-self-ion-simulations}, that the FES model produces comparable results to the EPH model. Nevertheless, it should be noted that the arbitrariness of the cutoff in the FES model is avoided in the EPH model. As for the effect of the interatomic potential in the accuracy of the results, from the mixing values reported in Table \ref{tab:mixingQ-self-ion-simulations} for the SW potential, it is expected that using the classical potentials can lead to a significant difference in the predicted mixing. In Ref. \cite{hamedani2021primary} it has been shown that among the GAP MLIP, Tersoff and SW potentials, the GAP potential exhibits the closest agreement with the experimental mixing values in Si.

\section{Conclusions}
\label{sec:conclusions}
In this paper we present new developments to TurboGAP, a versatile MD software specifically designed for computationally efficient simulations with GAP MLIPs. Accurate radiation damage simulations necessitate specific MD code capabilities, which are adaptive calculation of timestep, cooling the simulation cell at its boundaries, and accounting for atomic energy dissipation via electronic stopping. In this work, we detail the implementation of these three essential modules within TurboGAP, significantly enhancing its capability for radiation damage simulations.

To model electronic energy losses, two dissipation models have been implemented: the EPH model and the Friction-based electronic stopping, FES model. The EPH model, based on electron density-dependent coupling of electronic and atomic subsystems, was integrated into a two-temperature MD framework. The FES model, conversely, incorporates pre-calculated electronic stopping power as a friction term in conventional MD. Scaling tests confirmed that these implementations maintain the computational efficiency of the TurboGAP code.

For this study, a GAP MLIP was retrained for silicon (Si) using 2-body, 3-body, and turboSOAP descriptors on a general-purpose dataset. The ZBL repulsive potential was augmented to the GAP to accurately describe high-energy atomic repulsions during collision cascades. Our implementation was benchmarked against a similar plugin in LAMMPS, involving equilibration tests and comprehensive monitoring of energy loss and observables from both atomic and electronic subsystems.

Leveraging the trained MLIP and newly implemented modules, we conducted primary radiation damage simulations in Si, exploring PKA energies up to 10 keV and system sizes reaching 1 million atoms. We investigated the impact of various EPH model settings on the predicted defect state, specifically examining the effect of the size of the electronic subsystem, the spatial temperature gradient within it, and the temperature-dependence of electronic heat capacity and conductivity. The results indicate that these EPH model settings do not significantly influence the predicted defect state. There is a contrasting feature in the FES model, where the choice of kinetic energy cutoff does affect the finally observed defect state. Our cascade simulations reveal that the number of surviving defects and energy loss to the electronic system are affected by the chosen cutoff in the FES model. The dependence of the results on the ad-hoc selection of the kinetic energy cutoff in the FES model is removed in the EPH model, which imposes no such threshold. This emphasizes on the predictive power of the EPH model over the FES model, as the EPH model accurately describes relevant physics regimes without external constraints. Apart from the defect count and dissipated energy to the electronic system, the predicted atomic temperature and clustering of surviving defects differ between the models. The EPH model produces larger average defect clusters, with clusters up to ~180 defects, whereas the biggest clusters in the FES model contains fewer than 100 atoms.

Realistic accounting of electronic effects influences the evolution of the cascades as well, which is reflected in the ion-beam mixing values calculated from the simulations with the EPH model, showing good agreement with experimental values. Using Si as a case study, this work demonstrates that the simulation of radiation-induced damage in large atomic systems, leveraging GAP MLIPs and realistic dissipation models, will significantly benefit from the developments presented herein.

\section*{acknowledgment}
This work was partly supported by the Research Council of Finland through Project No. 349622, and by the European Union (ERC-2022-STG, project MUST, No. 101077454). AES acknowledges the financial support from Aaltonen. The authors acknowledge the computational resources provided by the Aalto Science-IT project and CSC—IT Center for Science, Finland. 
\\
\\
Views and opinions expressed herein are those of the authors only and do not necessarily reflect those of the European Union or the European Research Council Executive Agency. Neither the European Union nor the granting authority can be held responsible for them.

%

\end{document}